\documentclass[aps,prl,twocolumn,superscriptaddress,reprint]{revtex4-2}

\usepackage[utf8]{inputenc}
\usepackage[T1]{fontenc}
\usepackage[german,english]{babel}
\usepackage{amsmath}
\usepackage{mathtools}
\usepackage{amssymb}
\usepackage{bbm}
\usepackage{graphicx}
\usepackage{hyperref}
\usepackage[arrow, matrix, curve]{xy}
\usepackage{bm}
\usepackage{yhmath}
\usepackage{color}
\usepackage{url}
\usepackage{mathrsfs}
\usepackage{verbatim}
\usepackage{tabularx}
\usepackage{textcomp}
\usepackage{wrapfig}
\usepackage{enumitem}

\makeatletter
\renewcommand{\fnum@figure}{Fig. \thefigure}
\renewcommand{\fnum@table}{Tab. \thetable}

\renewcommand{\vec}[1]{\mathbf{#1}}
\renewcommand{\phi}[0]{\varphi}

\makeatother

\usepackage{braket}
\newcommand\diff{\mathrm{d}}

\renewcommand\vec[1]{\boldsymbol{\mathrm{#1}}}
\renewcommand\i{\text{i}}

\usepackage[usenames,dvipsnames]{xcolor}
\hypersetup{colorlinks=true, linkcolor=BrickRed, urlcolor=blue!50!black, citecolor=blue!50!black}

\usepackage[normalem]{ulem}

\makeatletter
\newcommand\hide@visible[1]{%
  \bgroup\fboxsep=.3ex\colorbox{Gray}{begin hide}%
  #1\colorbox{Gray}{end hide}\egroup%
}
\newcommand\hide@hidden[1]{%
  \bgroup\fboxsep=.3ex\colorbox{Gray}{hidden text}%
}
\newcommand\hide@invisible[1]{}
\newcommand\makevisible{\let\hide\hide@visible}
\newcommand\makehidden{\let\hide\hide@hidden}
\newcommand\makeinvisible{\let\hide\hide@invisible}
\makeatother
\makehidden

\begin{document}


\title{Resonant diffusion of a gravitactic circle swimmer}


\author{Oleksandr Chepizhko}
\author{Thomas Franosch}
\affiliation{Institut f\"{u}r Theoretische Physik, Universit\"{a}t Innsbruck,
Technikerstra\ss e 21A, A-6020, Innsbruck, Austria}


\date{\today}

\begin{abstract}
We investigate the dynamics of a single chiral active particle subject to an external  torque  
due to the presence of a gravitational field.  Our computer simulations reveal an arbitrarily strong increase of the  long-time diffusivity  of the gravitactic agent when  the  external torque approaches  the intrinsic angular drift. 
 We provide analytic expressions
 for  the mean-square displacement in terms of eigenfunctions and eigenvalues of the noisy-driven-pendulum problem. 
The pronounced maximum in the diffusivity is then rationalized by the vanishing of the lowest eigenvalues of the Fokker-Planck equation for the angular motion as the rotational diffusion decreases and  the underlying classical bifurcation is approached. 
A simple harmonic-oscillator picture for the barrier-dominated motion provides a quantitative description for the onset of the resonance while its range of validity is determined by the crossover to a critical-fluctuation-dominated regime.  
\end{abstract}


\maketitle

Active particles capable of self-propulsion by  converting energy into directed motion have come into 
research focus  and are important from both a  fundamental and  an applied point of view~\cite{Romanczuk2012,Elgeti2015,Bechinger2016,Zottl2016,Gompper2020}. 
Examples for active agents include various motile organisms, in particular bacteria~\cite{Berg1972, Berg1990,Lauga2006} or algae~\cite{Merchant2007}, as well as  artificial realizations such as Janus rods~\cite{Brown2016}, spheres~\cite{Buttinoni2012}, or Quincke rollers~\cite{Bricard2013}. 
Recently, significant advances in our understanding of transport properties of active motion  in homogeneous environments~\cite{Teeffelen2008,Sevilla2014,Kurzthaler2016,Zottl2016,Toner2016,Kurzthaler2017,Kurzthaler2018} and in media crowded with obstacles~\cite{Volpe2011,Chepizhko2013,Takagi2014,Zeitz2017,Reichhardt2014,Reichhardt2018JPCM,Chepizhko2019} have been achieved. 

External fields, torques and gradients  induce various forms of \emph{taxis} such as chemotaxis~\cite{Liebchen2018,Vuijk2021}, magnetotaxis~\cite{Erglis2007,Faivre2008}, gravitaxis~\cite{Roberts2006,Roberts2010}, rheotaxis~\cite{Mathijssen2019}, or viscotaxis~\cite{Liebchen2018b}
 due to a combination of the persistence of motion and different noise sources~\cite{Romanczuk2012,Zottl2016}.
Already the case of a homogeneous force field such as gravity  gives rise to counterintuitive dynamics by coupling to the orientational motion. 
For systems of $L$-shaped chiral microswimmers~\cite{tenHagen2014} and  
Janus rods~\cite{Brosseau2021}, mass-anisotropic colloids~\cite{Campbell2017,Singh2018}, and 
microorganisms~\cite{Roberts2006, *Roberts2010} gravitaxis has been demonstrated experimentally. In particular, these experiments have discovered  that chiral active particles subject to a gravitational field can move upwards, which was also supported by simulation studies in bottom-heavy microswimmers~\cite{Ruehle2020} and chiral swimmers~\cite{Fadda2020}. Experimental studies of sedimenting active particles revealed an enhancement of diffusivity by activity~\cite{Palacci2010}, while 
an analytical solution for the density profile   has been obtained only recently~\cite{Vachier2019}.
The sedimentation profile of active particles was also studied in detail in computer simulations and experiments with Janus particles~\cite{Ginot2018} as well as for run-and-tumble particles~\cite{Nash2010}.
Yet, the temporal dynamics and transport properties such as the mean-square displacement and the corresponding diffusivity have not been  elucidated in detail.


In this Letter, 
we demonstrate by computer simulation that gravitaxis of circle swimmers displays resonant diffusivities for small orientational diffusion as the graviational torque approaches   the intrinsic angular drift velocity. 
 We then 
elaborate a complete analytical solution of  gravitactic motion for the model developed by ten Hagen \emph{et al}~\cite{tenHagen2014}. A formal expression for the intermediate scattering function, encoding the spatio-temporal motion,  is derived. Using a time-dependent perturbative approach we extract the mean drift  and the mean-square displacement. Our analytic results  reveal that the resonance is encoded in the vanishing of the eigenvalues of the associated Fokker-Planck operator. We rationalize the onset of the resonance within a harmonic approximation and determine the  growth of  the maximum of the resonance by crossover scaling.

\paragraph*{Model.--} We rely on the model derived in Ref.~\cite{tenHagen2014} for an active chiral particle subject to an external  (gravitational) field. The particle moves at  constant speed $v$ along a direction $\vec{u}(t) :=(\cos \vartheta(t), \sin \vartheta(t))^T$ 
 parametrized by a time-dependent angle $\vartheta(t)$ measured from the horizontal 
\begin{equation}
\dot{\mathbf{r}}(t) = v\vec{u}(t) = v (\cos \vartheta(t), \sin \vartheta(t))\,.
\label{eq:r_dot}
\end{equation}
%
%
The evolution of $\vartheta(t)$ is governed by two contributions.
The external field results in an angle-dependent torque aligning the orientation in a certain direction. The coating of the active particle is such that with this orientation the self-propulsion is horizontal. Additionally, the anisotropy of the particle induces an internal angular drift $\omega>0$ resulting in the equation of motion  
\begin{equation}
\dot{\vartheta}(t) = \omega - \gamma \sin \vartheta(t) + \zeta(t)\,.
\label{eq:vartheta_dot}
\end{equation}
Here, the torque $\gamma$ is proportional to the external force and $\zeta(t)$ is a centered Gaussian white noise $\langle \zeta(t) \zeta(t') \rangle = 2 D_{\text{rot}} \delta(t-t')$,
where 
 $D_{\text{rot}}$ is the bare orientational diffusion constant.   
The equations of motion are rewritten with proper substitutions from Ref.~\cite{tenHagen2014}, where for simplicity we discard noise terms and additional drift terms due to (anisotropic) translational diffusion in Eq.~\eqref{eq:r_dot}. The neglected terms can be  incorporated with little effort but do not change the overall picture of the resonance phenomenon (see Supplemental Material~\cite{supplement_gravitaxis}).
%
%
Without the external field the model reduces to the free circle swimmer~\cite{Teeffelen2008,Friedrich2008,Loewen2016,Kurzthaler2017}, while for $\gamma > 0$ the angular motion  corresponds to Brownian motion in a tilted washboard potential \cite{Reimann2001, * Reimann2002, Lopez2020}. 
Ignoring the noise in Eq.~\eqref{eq:vartheta_dot} 
yields the classical dynamics of an  overdamped driven pendulum displaying a saddle-node bifurcation at the critical value $\gamma_c = \omega$~\cite{Strogatz2018nonlinear}. The mapping of the gravitaxis problem to the noisy driven pendulum constitutes our first result.

\paragraph*{Simulation.--} The model is encoded in 4 parameters characterizing gravitaxis.  We employ $1/\omega$ as  fundamental unit of time, while the radius of the circular motion $v /\omega$ sets the unit of length. Then $\gamma/\omega$ is the dimensionless  torque and $D_{\text{rot}}/\omega$ quantifies the relative importance of fluctuations.    
Stochastic simulations are performed and  the displacement $\Delta \vec{r}(t) := \vec{r}(t)-\vec{r}(0)$ is monitored in the stationary state. In particular, we extract the mean displacement $\langle \Delta \vec{r}(t) \rangle $  and the variance $\textsf{Var}[\Delta \vec{r}(t)] :=\langle [\Delta \vec{r}(t)- \langle \Delta \vec{r}(t) \rangle ]^2 \rangle$.

 In the stationary state the mean displacement grows linearly in time with the average velocity  $v (\langle \cos \vartheta(t) \rangle, \langle \sin \vartheta(t) \rangle )$. Since the stationary distribution of the angle $p^{\text{st}}(\vartheta)$ is elementary~\cite{Risken1989} the mean drift can be readily obtained by quadrature. Here, we recall that   
without noise, $D_{\text{rot}}=0$,  the orientational angle is locked at $0 < \vartheta_* \leq \pi/2$  with $\sin\vartheta_* = \omega/\gamma$ provided the  torque fulfills $\gamma \geq \omega$. 
If the  torque is weaker than the internal drift, $\gamma <  \omega$, the angular motion is periodic~\cite{Strogatz2018nonlinear}. Directly at the classical bifurcation, the particle moves 
upwards against the field~\cite{footnote1}. Upon reintroducing the noise the average horizontal motion  is suppressed above the bifurcation, while below the fluctuations enable a net drift (see also Fig.~S.3 in SM~\cite{supplement_gravitaxis}). The drift against the field direction is always suppressed by noise.

\begin{figure}[!ht]
\includegraphics[width=\linewidth]{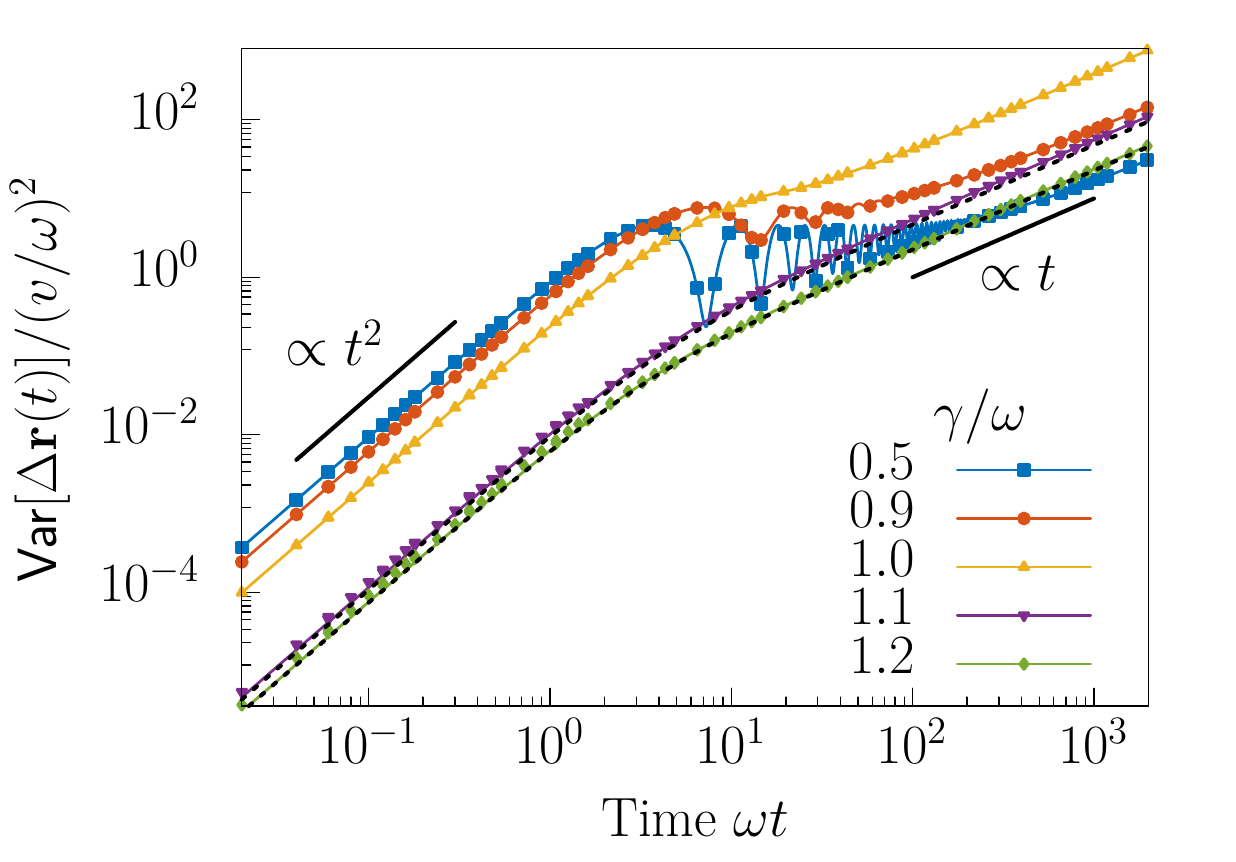}%
\caption{Time-dependent variance $\textsf{Var}[\Delta \vec{r}(t)]$ on logarithmic scales for a rotational diffusivity of $D_{\text{rot}}/\omega=0.005$\footnote{Unfortunately, the published version states a wrong value.} for  
different torques $\gamma$. Symbols correspond to simulation, the solid lines represent the analytic solution. 
The dotted black lines are  evaluated  within the harmonic oscillator approximation, Eq.~\eqref{eq:harmonic_diffusivity}, for $\gamma=1.1 \omega$ and $\gamma=1.2\omega$. 
The thick black lines are power laws serving as guide to the eye. 
}
 \label{fig:Fig1}
\end{figure}

\begin{figure}[!h]
\includegraphics[width=\linewidth]{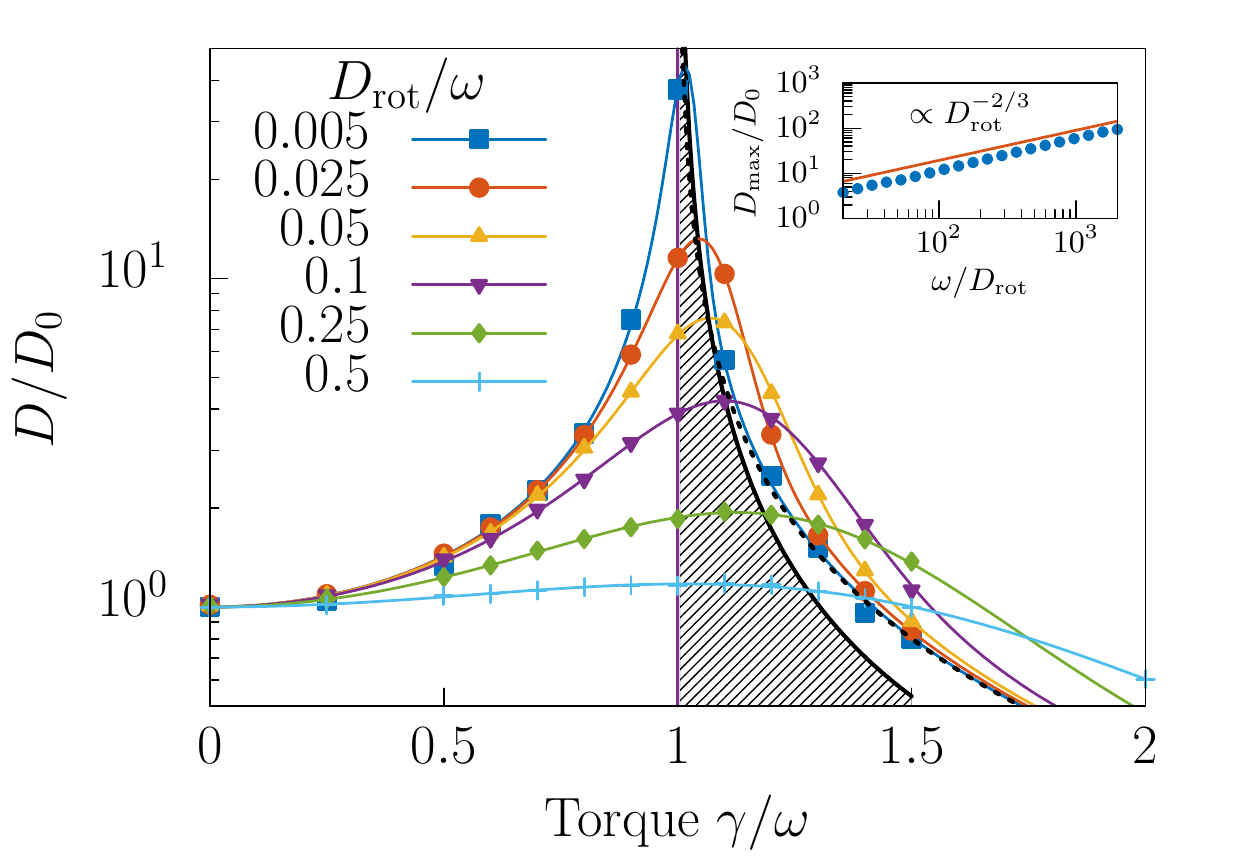}%
\caption{Resonance of diffusivity.  The diffusivity $D$ 
as a function of the torque $\gamma$ for different rotational diffusivities $D_{\text{rot}}$. The diffusivity is normalized to the diffusion coefficient of a free circle swimmer $D_0 = v^2 D_{\text{rot}} /[2 (D_{\text{rot}}^2 + \omega^2)]$. Symbols correspond to simulation, the full lines are analytical results. 
The black dotted line corresponds to the harmonic oscillator approximation, Eq.~\eqref{eq:harmonic_diffusivity}, for $D_{\text{rot}} = 0.005 \omega$. 
The right boundary of the shaded area corresponds to the parametric curve $\hat{D}_{\text{rot}} :=   |\gamma/\omega-1|^{-3/2} D_{\text{rot}}/\omega =\text{const.}$ asymptotically intersecting the maxima of the diffusion coefficients.  
Inset: Increase of the  maximal diffusion coefficient $D_{\text{max}}$ for increasing inverse noise $1/D_{\text{rot}}$ 
 on logarithmic scales obtained from theory.\footnote{Unfortunately, this inset contains wrong tic marks in the published version.} 
}
\label{fig:Fig2}
\end{figure}

The variance $\textsf{Var}[\Delta \vec{r}(t)]$ increases as $t^2$ for small times, see Fig.~\ref{fig:Fig1}, where the prefactor decreases drastically as the torque is increased. Below the classical bifurcation characteristic oscillations emerge, similar to the free circle swimmer
\cite{Kummel2013,Kurzthaler2017},
while for $\gamma> \omega$ the variance increases monotonically. The long-time behavior is diffusive in the presence of noise, $D_{\text{rot}}>0$, and the diffusion coefficient is enhanced close to the classical bifurcation. The extracted long-time diffusion coefficients $D := \lim_{t\to \infty}(1/4) \diff \textsf{Var}[\Delta \vec{r}(t)] /\diff t  $ are displayed in terms of the known diffusivity without external torque ~\cite{Teeffelen2008,Kurzthaler2017}  $D_0=v_0^2 D_{\text{rot}}/[2(D_{\text{rot}}^2+\omega^2)]$
  in  Fig.~\ref{fig:Fig2}.  
Close to the classical bifurcation a resonance emerges that becomes narrower and more pronounced as the noise is decreased.

\paragraph*{Theory.--} The goal of this part is to provide a theoretical explanation for the found resonance and to derive scaling laws in the vicinity of the classical bifurcation. 
The dynamical properties are encoded in the propagator $\mathbb{P}( \vec{r}, \vartheta, t | \vartheta_0)$, i.e.\@ the  
conditional probability distribution that the particle has displaced by $\vec{r}$ and its instantaneous speed exhibits an orientation $\vartheta$ at lag time $t$ given it started with orientation $\vartheta_0$. Angles are considered to 
be $2\pi$-periodic. 
We focus on its spatial Fourier transform $\tilde{\mathbb{P}} := \tilde{\mathbb{P}}(\vec{k},\vartheta,t | \vartheta_0)=\int d \mathbf{r} \exp(- \mathrm{i}\vec{k} \cdot \vec{r}) \mathbb{P}( \vec{r}, \vartheta, t | \vartheta_0)$
and derive by standard methods~\cite{Risken1989} the (Fourier transformed) Fokker-Planck equation  (see also SM~\cite{supplement_gravitaxis})
\begin{align}\label{eq:eom}
\partial_t \tilde{\mathbb{P}} &= - \partial_\vartheta [ (\omega - \gamma \sin \vartheta) \tilde{\mathbb{P}} ] 
+ D_{\text{rot}} \partial_\vartheta^2 \tilde{\mathbb{P}} - i v  \vec{k} \cdot  \vec{u} \tilde{\mathbb{P}} \nonumber \\
 &=: (\mathcal{L} + \delta \mathcal{L}_{\vec{k}} ) \tilde{\mathbb{P}}. 
\end{align}
Here the  operator $\mathcal{L}$ encodes the motion of the angle, while  $\delta \mathcal{L}_{\vec{k}} = - i v \vec{k} \cdot \vec{u} $ describes the coupling to the translational dynamics. 
The formal solution 
is thus $\tilde{\mathbb{P}}(\vec{k},\vartheta, t) |\vartheta_0) = \exp[(\mathcal{L} + \delta \mathcal{L}_{\vec{k}} )t] \delta(\vartheta-\vartheta_0)$.  
 
From this quantity the intermediate scattering function (ISF) $F(\mathbf{k},t)=\langle \exp(-\i \vec{k} \cdot \Delta \vec{r}(t)) \rangle$ is obtained in the stationary state by averaging over the initial angle and integrating over the final one   
\begin{align}\label{eq:ISF_definition}
F(\vec{k},t)  &= \int_0^{2\pi}\!\! \diff \vartheta\!\! \int_0^{2\pi} \! \!
\diff \vartheta_0 \, \tilde{\mathbb{P}}(\vec{k},\vartheta, t | \vartheta_0) p^{\text{st}}(\vartheta_0) \nonumber \\
&=  \int_0^{2\pi}\!\! \diff \vartheta   \exp[(\mathcal{L} + \delta \mathcal{L}_{\vec{k}} )t] p^{\text{st}}(\vartheta)
.
\end{align}
Moments of the displacement can be extracted by a series expansion in the wave vector $\vec{k}$
\begin{equation}\label{eq:ISF_series}
F(\vec{k}, t) = 1 - \mathrm{i} \vec{k} \cdot \langle \Delta \vec{r} (t)\rangle -\frac{1}{2} \langle [\vec{k} \cdot \Delta \vec{r}(t)]^2 \rangle +\ldots\,.
\end{equation}

Here  we follow the strategy of Ref.~\cite{Kurzthaler2017} (see also Ref.~\cite{Lapolla2020}),  solve the eigenvalue problem of the reference system  $\mathcal{L}$, and apply time-dependent perturbation theory for $\delta \mathcal{L}_{\vec{k}}$. 
We denote by $\{ | n \rangle : n\in \mathbb{Z} \}$  the standard orthonormal basis   in the Hilbert space $L^2[0,2\pi]$ with real-space representation $\langle \vartheta | n \rangle := \exp(i n \vartheta)/ \sqrt{2\pi}$. In this basis $\mathcal{L}$ is represented in terms of its matrix elements~\cite{Risken1989} 
\begin{align}\label{eq:matrix_elements}
\mathcal{L}_{mn} &= \langle m | \mathcal{L}  n \rangle := \int_0^{2\pi} \frac{\diff \vartheta}{2\pi} e^{-i m \vartheta } \mathcal{L} e^{in \vartheta}  \nonumber \\
&= (-D_{\text{rot}} m^2 - i m \omega ) \delta_{mn} + \frac{\gamma}{2} m ( \delta_{m,n+1} - \delta_{m,n-1} ) . 
\end{align}
In particular, the matrix representation is non-Hermitian and tridiagonal due to the torque. 
Right and left eigenstates $ \mathcal{L} | r_\lambda \rangle = -\lambda | r_\lambda \rangle, \mathcal{L}^\dagger | l_\lambda \rangle = -\lambda^* |l_\lambda\rangle $ are readily obtained by numerically diagonalizing the matrix, Eq.~\eqref{eq:matrix_elements}, yielding the expansion coefficients $\langle n | r_\lambda \rangle$, $ \langle l_\lambda | n \rangle$ as right and left eigenvector of the matrix $\mathcal{L}_{mn}$. 
The corresponding real-space representation is obtained by expansion $r_\lambda(\vartheta) = \sum_{n\in \mathbb{Z}} \langle \vartheta | n \rangle \langle n | r_\lambda \rangle $. 
By conservation of probability $\lambda=0$ is an eigenvalue and the real-space representation of the associated eigenstates are
$ r_0(\vartheta) = p^{\text{st}}(\vartheta)$ and $l_0(\vartheta) = 1$.  Comparing with Eq.~\eqref{eq:ISF_definition}
   we find the compact expression for the ISF 
\begin{align}
F(\vec{k},t) &= \langle l_0 |  \exp[(\mathcal{L} + \delta \mathcal{L}_{\vec{k}} )t] r_0 \rangle . 
\end{align}
Since $\delta \mathcal{L}_{\vec{k}} = -i v\vec{k} \cdot \vec{u}$, 
time-dependent perturbation theory (Born series) \cite{Sakurai2011modern}   
\begin{align}
e^{(\mathcal{L}+\delta \mathcal{L}_{\vec{k}} ) t } =& e^{\mathcal{L} t} + \int_0^t \diff s\, e^{\mathcal{L} (t-s)} \delta \mathcal{L}_{\vec{k}}  e^{\mathcal{L} s } \nonumber \\
& + \int_0^t \diff s \int_0^s \diff u \,  e^{\mathcal{L} (t-s)} \delta \mathcal{L}_{\vec{k}} e^{\mathcal{L} (s-u)} \delta \mathcal{L}_{\vec{k}}  e^{\mathcal{L} u } \nonumber \\
& + O(\delta \mathcal{L}_{\vec{k}})^3 ,
\end{align} 
directly yields the moments of the series expansion in $\vec{k}$ as in Eq.~\eqref{eq:ISF_series}. 
In particular, 
expanding to second order in $\vec{k}$ and squeezing in the completeness relation $\sum_\lambda | r_\lambda \rangle\langle l_\lambda|  = \mathbbm{1}$, the time integrals can be formally performed (see also SM~\cite{supplement_gravitaxis}). Then  
 we read off the mean drift velocity and the variance along the direction $\vec{n} := \vec{k} /k $ 
\begin{subequations}\label{eq:moments_analytics}
\begin{align}
\vec{n} \cdot  \frac{\diff}{\diff t} \langle \Delta \vec{r}(t) \rangle  &= \frac{i}{k} \langle l_0 | \delta \mathcal{L}_{\vec{k}}   r_0 \rangle , \\
\textsf{Var}[\vec{n}\cdot \Delta \vec{r}(t)]
&= \frac{2}{k^2}  \sum_{\lambda\neq 0} \frac{ 1- \lambda t- e^{-\lambda t}}{\lambda^2} 
 \langle l_0 |\delta \mathcal{L}_{\vec{k}}   r_\lambda \rangle \langle l_\lambda |  \delta \mathcal{L}_{\vec{k}}   r_0 \rangle , \label{eq:MSD_analytics}
 \end{align}
\end{subequations}
and infer the associated long-time diffusion coefficient
\begin{align}\label{eq:Diffusion_analytics}
D_{\vec{n}}   &= \frac{-1}{k^2}  \sum_{\lambda\neq 0} \frac{1}{\lambda} 
 \langle l_0 |\delta \mathcal{L}_{\vec{k}}   r_\lambda \rangle \langle l_\lambda |  \delta \mathcal{L}_{\vec{k}}   r_0 \rangle . 
 \end{align}
The analytical expressions 
for  the total variances as well as the corresponding diffusion coefficient   perfectly match the simulations, see  Figs.~\ref{fig:Fig1} and \ref{fig:Fig2}. 
The exact expressions in terms of eigenfunctions and eigenvalues suggest that at resonance the eigenvalues become smaller and smaller as the rotational diffusion coefficient is decreased, which is confirmed by numerical diagonalization (see SM~\cite{supplement_gravitaxis}).

To gain further analytical insight, we evaluate Eqs.~\eqref{eq:moments_analytics} and~\eqref{eq:Diffusion_analytics} within   
 a harmonic approximation for the motion close to the classical fixed point $\vartheta_*$. 
For small $D_{\text{rot}} \ll \omega$ and well above the bifurcation the fluctuations are anticipated to be small and the corresponding linearized Langevin equation reads 
\begin{align}
\dot{\vartheta}(t) = - \frac{1}{\tau}[ \vartheta(t) - \vartheta_*] + \zeta(t) ,
\end{align} 
with vanishing relaxation rate $1/\tau = \sqrt{\gamma^2-\omega^2} \to 0$ as $\gamma \downarrow \omega$. The associated eigenvalues of the overdamped harmonic oscillator are then simply $\lambda_n = n /\tau, n \in \mathbb{N}_0$~\cite{Risken1989}, and indeed they approach zero for $\gamma \downarrow 0$ (see also SM~\cite{supplement_gravitaxis}). 

To leading order we replace the perturbing operator $\delta \mathcal{L}_{\vec{k}}$ by the complex number  $i v \vec{k} \cdot \vec{u}_*$ with the fixed orientation $\vec{u}_* = (\cos\vartheta_*, \sin \vartheta_*)$. In this approximation, the drift velocity,  is non-fluctuating and assumes its classical value. In the same approximation the variance and diffusion coefficient vanish by orthogonality of the eigenstates, consistent with a purely deterministic and non-chaotic motion. To leading non-trivial order we replace 
\begin{align}
\delta \mathcal{L}_{\vec{k}} \simeq - i v \vec{k} \cdot \vec{u}_* + (\vartheta-\vartheta_*) \left. \frac{\partial}{\partial \vartheta} \delta \mathcal{L}_{\vec{k}} \right|_{\vartheta=\vartheta_*} .
\end{align}
Then within the harmonic oscillator approximation, also the off-diagonal matrix elements can be evaluated analytically, in particular, their magnitude is proportional to the angular oscillator width $\langle (\vartheta-\vartheta_*)^2 \rangle = \sqrt{D_{\text{rot}} \tau}$ 
(see  SM~\cite{supplement_gravitaxis} for details). Furthermore, the angular position operator $\vartheta-\vartheta_*$ induces 
 only non-vanishing transition matrix element in Eqs.~\eqref{eq:MSD_analytics},\eqref{eq:Diffusion_analytics}  coupling the ground state to the first excited state. Therefore the  sums reduce to a single term and we find the compact expressions
\begin{subequations}
\begin{align}
\textsf{Var}[\vec{n}\cdot \Delta \vec{r}(t)] 
 \simeq & \, 2 D_{\vec{n}} \tau  \left( \frac{t}{\tau} -1+  e^{-t/\tau} \right), \\
D_{\vec{n}}   \simeq & \, (v \tau)^2  D_{\text{rot}}  (n_x \sin \vartheta_* - n_y \cos\vartheta_*)^2 .
\label{eq:harmonic_diffusivity}
\end{align}
\end{subequations}
Within the harmonic oscillator approximation the variance is strictly proportional to $D_{\text{rot}}$ suggesting that for sufficiently small orientational fluctuations the curves in Figs. ~\ref{fig:Fig1} and \ref{fig:Fig2} should approach a master curve.
The corresponding curves  are included in Figs.~\ref{fig:Fig1} and \ref{fig:Fig2} as black dotted lines and are in  quantitative agreement for small $D_{\text{rot}} \ll \omega$ and  $\gamma > \omega$ not too close to the bifurcation.  
The emergence of the resonance is thus rationalized in terms of the softening of the harmonic relaxation rate $1/\tau \to 0$ as the  
bifurcation is approached. 

The harmonic picture suggests that the diffusion coefficient becomes infinite directly at the bifurcation while the simulation and the full analytic expression predict a rounding with a maximal diffusivity. 
The picture of the harmonic oscillator should hold  provided the barrier is sufficiently high such that Kramers' escape rate~\cite{Risken1989} is much smaller than the harmonic relaxation rate. In terms of the effective potential $ D_{\text{rot}} U(\vartheta) /k_B T = - \omega \vartheta - \gamma \cos \vartheta$ the barrier height $\Delta U := U(\pi-\vartheta_*) - U(\vartheta_*)$ reduces to 
\begin{align}
\frac{\Delta U}{k_B T} = \frac{4\sqrt{2}}{3}\frac{\omega}{D_{\text{rot}}} \epsilon^{3/2} [1 + O(\epsilon) ] ,
\end{align}
where 
we introduced the separation parameter $\epsilon := (\gamma - \omega )/\omega$ for the distance to the bifurcation (see  SM~\cite{supplement_gravitaxis} for details). This observation suggests introducing the reduced rotational diffusion coefficient $\hat{D}_{\text{rot}} := |\epsilon|^{-3/2} D_{\text{rot}} /\omega$ such that for $\hat{D}_{\text{rot}} \ll 1, \epsilon>0$ the harmonic approximation holds, while for $\hat{D}_{\text{rot}} \gg 1$ the barrier can be crossed readily by fluctuations and the precise height of the barrier should be irrelevant.
By matching the critical fluctuation to the harmonic oscillator we predict that the maximal diffusivity should occur at $\hat{D}_{\text{rot}} = O(1)$ or $D_{\text{rot}} \propto |\epsilon|^{3/2}$. For small $\epsilon>$, the 
relaxation time diverges as $\tau \propto |\epsilon|^{-1/2}$
and  from Eq.~\eqref{eq:harmonic_diffusivity}
we infer   the scaling law for the maximal diffusivity
\begin{align}
D_{\text{max}}/D_0   \propto |\epsilon|^{-1} \propto D_{\text{rot}}^{-2/3} \qquad \text{for } D_{\text{rot}} \to 0, 
\end{align}
where the we used that $D_0 \propto D_{\text{rot}}$ as $D_{\text{rot}}\to 0$.  
The numerical values nicely follow the prediction  asymptotically as shown  in Fig.~\ref{fig:Fig2}.

\paragraph*{Summary and Conclusion.--}
We have demonstrated that the long-time translational diffusivity of a chiral active Brownian particle displays a resonance for the external torque approaching the intrinsic angular drift. The resonance originates from an underlying bifurcation of the classical driven pendulum. There are certain similarities 
with the giant diffusion in  tilted washboard potentials~\cite{Reimann2001, * Reimann2002}, yet our approach 
of decomposition into eigenfunctions is rather complementary and allows calculating the entire time-dependence of low-order moments and is in spirit closer to Ref.~\cite{Lopez2020}. The connection of the resonance to the vanishing low-lying eigenvalues is uncovered and an intuitive picture in terms of a competition between barrier-dominated and critical-fluctuation-dominated regime is developed. 

The harmonic approximation is surprisingly accurate for the variance as well as for the diffusivity, despite ignoring the rare activation processes over the barrier. We conclude that far above the bifurcation, the orientation performs only small fluctuations most of the time close to the minimum of the effective potential. Then the diffusion coefficient for the translational diffusion remains also small. Yet, approaching the bifurcation these fluctuations become more significant as the confining angular potentials becomes softer yielding a significant enhancement of the diffusivity. For too large angular fluctuations the harmonic approximation breaks down as barrier-crossing events become important. After such barrier crossings the orientation quickly completes a full turn until getting stuck again. From the accurateness of the harmonic oscillator description we conclude that these fast events do not significantly contribute to the diffusivity. The essence of the resonance is thus due to the enhancement of small fluctuations as provided by the orientational diffusion coefficient exploring a softening potential as the bifurcation is approached.

Our theoretical work makes detailed predictions for active motion of chiral agents in external fields that can be tested in experiments, both for artificial and biological asymmetric microswimmers. 
The method is readily extended to calculate higher moments such as the skewness, the non-Gaussian parameter, or the complete intermediate scattering function. 
 While we explicitly considered gravitaxis, the underlying equations are rather generic and the analysis and methods presented should  transfer with suitable adjustments to other forms of \emph{taxis} at the microscale  such as durotaxis~\cite{Lo2000}, chemotaxis~\cite{Vuijk2021}, thermotaxis~\cite{Auschra2021}, or topotaxis~\cite{Schakenraad2020}. Also, the results could be valuable for the transport rigid polymer solutions in a flow near a wall~\cite{Park2007}.

The resonance   is of interest   not only for single-particle transport in  external fields, but also has implications for the collective motion of chiral active particles with alignment interactions~\cite{Liebchen2017,Levis2018,Ventejou2021,Lei2019}  provided effective mean-field equations can be derived~\cite{Chepizhko2010,Peruani2008,Pimentel2008,Bolley2012} which would be similar in structure to the equations of motion studied here. 

Last our work has theoretical ramifications on the interplay of the critical slowing down of classical transport close to bifurcations and their smearing by random fluctuations. The evolution of the eigenspectrum also for other bifurcations, including the pitchfork or transcritical bifurcations,  should be relevant for various branches of science, such as the physics of the Josephson junction~\cite{Strogatz2018nonlinear} or collective (Kuramoto) synchronization~\cite{Acebron2005}.    

\nocite{Felderhof2014}
\nocite{Kuemmel2014reply}

\begin{acknowledgements}
\paragraph*{Acknowledgments.--}
We thank Christina Kurzthaler for constructive criticism on the manuscript.
OC is supported by the Austrian Science Fund (FWF): M 2450-NBL. TF acknowledges funding by FWF: P 35580-N.
 The computational results presented have
been achieved in part using the HPC infrastructure LEO of the University of Innsbruck.
\end{acknowledgements}

\clearpage
\onecolumngrid

\begin{center}
    {\bf SUPPLEMENTAL MATERIAL}
\end{center}

\vspace{0.25cm}

\setcounter{equation}{0}
\renewcommand*{\theequation}{SM\arabic{equation}}

\setcounter{figure}{0}
\renewcommand*{\thefigure}{SM\arabic{figure}}

\section{Simplifying the equations of motion and mapping to the noisy driven pendulum}

The stochastic equations of motion for an asymmetric circle swimmer in the presence of an external (gravitational) field were proposed by ten Hagen \emph{et al.} in  Ref.
~[1]. 
The goal of this section is to  show that the model greatly simplifies upon a simple rotation to the principal axis of the system. Furthermore the essence of gravitaxis can be discussed by discarding additional terms connected to translational diffusion and the associated noise. In this approximation the model reduces to the paradigmatic noise driven pendulum. Explicitly the equations of motion of Ref.~[1] 
read 
\begin{wrapfigure}{r}{.35\linewidth}
\centering
\includegraphics[width=0.8\linewidth]{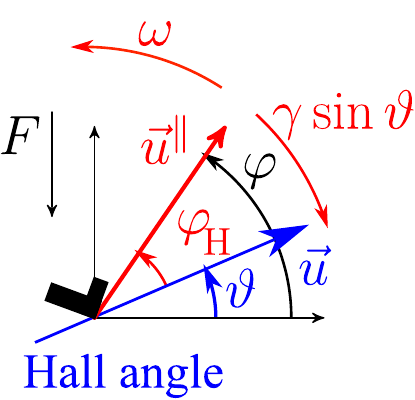}%
\caption{a) Illustration of the coordinate system used. The orientation of the principal axis of the particle is along $\vec{u}^\parallel =  (\cos\varphi, \sin\varphi)$, the drift velocity $\vec{u} = (\cos\vartheta, \sin\vartheta)$ is tilted by the Hall angle $\varphi_H$ to the particle's principal axes. The torque $\varphi_H$ tends 
to align the particle's direction with the Hall direction while the intrinsic angular drift rotates the direction counter-clockwise.  }
\label{fig:Hall_angle}
\end{wrapfigure}
\begin{subequations}
\begin{align}
\dot{\vec{r}}(t) &= \frac{P^*}{b} [\textsf{D}_T(t) \cdot \vec{u}^\perp(t) + \ell \vec{D}_C(t) ] + \beta \textsf{D}_T(t) \cdot \vec{F} + \boldsymbol{\zeta}_{\vec{r}}(t), \\
\dot{\varphi}(t) &= \frac{P^*}{b} [\ell D_{\text{rot}}+ \vec{D}_C(t) \cdot \vec{u}^\perp(t)] + \beta \vec{D}_C(t) \cdot \vec{F} + \zeta(t),
\end{align}
\end{subequations}
for the time-dependent  position $\vec{r}(t)$ and the orientational angle $\varphi(t)$ of the particle. 
The two unit vectors $\vec{u}^\parallel(t) = (\cos \varphi(t) , \sin \varphi(t))$ and $\vec{u}^\perp(t) = (-\sin \varphi(t), \cos\varphi(t))$ are associated with the orientational angle $\varphi(t)$, see Fig.~\ref{fig:Hall_angle} for illustration. Furthermore, $\vec{F} = (0,- m g \sin\alpha) = (0,-F) $ denotes the gravitational force, $\textsf{D}_T(t) = D_\parallel \vec{u}^\parallel(t) \otimes \vec{u}^\parallel(t) + D^\perp_\parallel [\vec{u}^\parallel \otimes \vec{u}^\perp(t) + \vec{u}^\perp(t) \otimes \vec{u}^\parallel(t)] + D_\perp \vec{u}_\perp(t) \otimes \vec{u}_\perp(t) $, 
$\vec{D}_C(t) = D_C^\parallel \vec{u}^\parallel(t) + D_C^\perp \vec{u}^\perp(t)$ encodes the translational diffusion and the translation-rotation coupling[2], $\beta =1/k_B T$ refers to the inverse (effective) temperature, $\ell$ and $b$ are characteristic lengths, and $P^*$ is the strength of the self propulsion.
Last, $\boldsymbol{\zeta}_{\vec{r}}, \zeta_{\phi}$ are independent centered Gaussian white noises of 
variance $ \langle \boldsymbol{\zeta}_{\vec{r}}(t) \otimes \boldsymbol{\zeta}_{\vec{r}}(t') \rangle = 2 \textsf{D}_T(t) \delta(t-t'), \langle \zeta(t) \zeta(t') \rangle = 2 D_{\text{rot}} \delta(t-t')$ respectively. 


Without restriction we assume the unit vectors $\vec{u}^\parallel(t), \vec{u}^\perp(t)$ to be chosen as the principal  axes of the  the tensor $\underline{\textbf{D}}_T$. In this frame the tensor becomes diagonal and  we henceforth ignore the couplings $D^\perp_\parallel$ and refer to the orientation of the particle as the angle of the principal axis of the larger eigenvalue $D^\parallel$ with respect to some space-fixed reference axis.

Evaluating the scalar products in the angular equation of motion yields
\begin{align}
\dot{\varphi}(t) &= \frac{P^*}{b}(\ell D_{\text{rot}} + D_C^\perp) - \beta F( D^\parallel_C \cos\varphi(t) + D^\perp_C \sin \varphi(t) ) + \zeta(t).
\end{align} 
 Using trigonometric addition formulas this simplifies to 
\begin{align}
\dot{\varphi}(t) &= \omega - \gamma \sin(\varphi(t)-\varphi_H) +   \zeta(t),
\end{align}
with the  angular drift velocity $\omega$, angular driving $\gamma$, and the \emph{Hall angle} $\varphi_H$ given by
\begin{subequations}
\begin{align}
\omega &= \frac{P^*}{b} (\ell D_{\text{rot}} + D_C^\perp), \qquad 
\gamma = \beta F \sqrt{ (D_C^\parallel)^2 + (D_C^\perp)^2 }, \\
\sin \varphi_H &= -\beta F D_C^\perp  /\gamma, \qquad  \cos \varphi_H =  \beta F D_C^\parallel /\gamma . 
\end{align}
\end{subequations}

Similarly the translational equation of motion simplifies to
\begin{align}\label{eq:Langevin_full}
\dot{\vec{r}}(t) &= \frac{P^*}{b} [ D_\perp \vec{u}^\perp(t) +\ell  D_C^\parallel \vec{u}^\parallel(t) + \ell D_C^\perp \vec{u}^\perp(t) ]  
+   \beta   D_\parallel [ \vec{F} \cdot \vec{u}^\parallel(t) ]  \vec{u}^\parallel(t) \cos \varphi(t) 
  +  \beta   D_\perp [\vec{F} \cdot \vec{u}^\parallel(t) ]  \vec{u}^\perp(t) \cos \varphi(t) 
+ \boldsymbol{\zeta}_{\vec{r}} 
\nonumber \\
&=: v \begin{pmatrix}
\cos (\varphi(t)- \varphi_H) \\
\sin (\varphi(t) - \varphi_H ) 
\end{pmatrix}  
+ v_1 \begin{pmatrix} 
-\sin \varphi(t) \\
\cos\varphi(t) 
\end{pmatrix}- v_2 \begin{pmatrix}
 \sin (2\varphi(t)) \\
 \cos(2 \varphi(t) ) 
\end{pmatrix}
- v_3 \begin{pmatrix}
0 \\
1  
\end{pmatrix} 
+
\boldsymbol{\zeta}_{\vec{r}} \nonumber \\
& =: \vec{V}(t)  + \boldsymbol{\zeta}_{\vec{r}}(t) ,
\end{align}  
with $v = P^* \ell \gamma / b \beta F$, $ v_1  = P^* D_\perp /b$, $v_2 = \beta F (D_\parallel- D_\perp)/2 $, $v_3 = \beta F (D_\parallel + D_\perp )/2$. 
Without circular motion and ignoring both rotational as well as translational thermal fluctuations, a stationary solution with $\varphi(t)= \varphi_H$ exists, the particle moves with constant velocity perpendicularly to the external field, see Fig.~\ref{fig:Hall_angle}.  We abbreviate 
 $\vartheta(t) = \varphi(t) - \varphi_H$ and find the e.o.m. 
\begin{align}
\dot{\vartheta}(t) &= \omega - \gamma \sin \vartheta(t)  +   \zeta(t) , \qquad 
\dot{\vec{r}}(t) =  
\vec{V}(t)  + \boldsymbol{\zeta}_{\vec{r}}(t)
,
\end{align} 
with independent centered Gaussian white noise characterized by $\langle \zeta(t) \zeta(t') \rangle =  2 D_{\text{rot}} \delta(t-t')$, $\langle \boldsymbol{\zeta}_{\vec{r}}(t) \otimes \boldsymbol{\zeta}_{\vec{r}}(t') = 2 \textsf{D}_T(t) \delta(t-t')$. Ignoring the terms induced by translational diffusion, the e.o.m.\@ for the translational motion simplifies to the one discussed in the main text
\begin{align}
\dot{\vartheta}(t) &= \omega - \gamma \sin \vartheta(t)  +   \zeta(t) , \qquad \dot{\vec{r}}(t) = v \vec{u}(t) := v 
\begin{pmatrix} 
\cos \vartheta(t) \\
\sin \vartheta(t) 
\end{pmatrix}
.
\end{align}

\section{Fokker-Planck equation, intermediate scattering function and perturbation theory}

\subsection{Fokker-Planck equation} 
By standard techniques [3] one derives the Fokker-Planck equation for the conditional probability density $\mathbb{P} = \mathbb{P}(\vec{r},\vartheta, t| \vartheta_0)$ to find the particle at speed orientation $\vartheta$ and displaced by $\vec{r}$ at time $t$ given the orientation at time $t=0$ is prescribed to $\vartheta_0$:
\begin{align}
\partial_t \mathbb{P} = - \partial_\vartheta [ (\omega - \gamma \sin \vartheta) \mathbb{P} ] +
 D_{\text{rot}} \partial_\vartheta^2 \mathbb{P} - \frac{\partial}{\partial \vec{r}} \cdot ( \vec{V} \mathbb{P} ) + \frac{\partial}{\partial  \vec{r}} \cdot \left( \textsf{D}_T \cdot \frac{\partial}{\partial  \vec{r}}  \mathbb{P} \right) .
\end{align}
The initial condition is provided by  
$\mathbb{P}(\vec{r},\vartheta, t=0 | \vartheta_0) = \delta(\vec{r}) \delta(\vartheta-\vartheta_0)$.

We also introduce the spatial Fourier transform 
\begin{align}
\tilde{\mathbb{P}}(\vec{k}, \vartheta, t | \vartheta_0) = \int \diff^2 r \exp(-i \vec{k} \cdot  \vec{r} ) \mathbb{P}(\vec{r},\vartheta,t| \vartheta_0),
\end{align}
with corresponding e.o.m.
\begin{align}\label{eq:eom}
\partial_t \tilde{\mathbb{P}} &= - \partial_\vartheta [ (\omega - \gamma \sin \vartheta) \tilde{\mathbb{P}} ] 
+ D_{\text{rot}} \partial_\vartheta^2 \tilde{\mathbb{P}} - i \vec{k} \cdot  \vec{V} \tilde{\mathbb{P}} - D_\perp k^2 \tilde{\mathbb{P}} - ( D_\parallel - D_\perp)(\vec{u}^\parallel \cdot \vec{k})^2 \tilde{\mathbb{P}}  \nonumber \\
&=: ( \mathcal{L} + \delta \mathcal{L}_{\vec{k}}) \tilde{\mathbb{P}}.
\end{align}
Here the  operator $\mathcal{L}$ encodes the motion of the angle, while  $\delta \mathcal{L}_{\vec{k}} = - i  \vec{k} \cdot \vec{V} - D_\perp k^2 - (D_\parallel - D_\perp ) (\vec{u} ^\parallel \cdot \vec{k} )^2 $ describes the coupling to the translational dynamics. 
The formal solution 
is thus $\tilde{\mathbb{P}}(\vec{k},\vartheta, t) |\vartheta_0) = \exp[(\mathcal{L} + \delta \mathcal{L}_{\vec{k}} )t] \delta(\vartheta-\vartheta_0)$.  
From this quantity the intermediate scattering function (ISF) $F(\mathbf{k},t)=\langle \exp(-i \vec{k} \cdot \Delta \vec{r}(t)) \rangle$ is obtained in the stationary state by averaging over the initial angle and summing over the final one   
\begin{align}\label{eq:ISF_definition}
F(\vec{k},t)  &= \int_0^{2\pi}\!\! \diff \vartheta\!\! \int_0^{2\pi} \! \!
\diff \vartheta_0 \, \tilde{\mathbb{P}}(\vec{k},\vartheta, t | \vartheta_0) p^{\text{st}}(\vartheta_0) =  \int_0^{2\pi}\!\! \diff \vartheta   \exp[(\mathcal{L} + \delta \mathcal{L}_{\vec{k}} )t] p^{\text{st}}(\vartheta)
,
\end{align}
where $p^{\text{st}}(.)$ is the non-trivial distribution of the orientation in the steady state.  We will derive low-order spatial moments 
by series expansion in powers of the wavevector $\vec{k}$.

It is favorable to  rely on  the isomorphism of periodic square-integrable functions $f(\vartheta)\in L^2[0,2\pi]$ and abstract states $| f \rangle \in \mathcal{H}$ [4]. 
The relation is made explicit by introducing generalized eigenstates $|\vartheta \rangle$ such that $
f(\vartheta) = \langle \vartheta | f \rangle$. For this to hold we require 
\begin{align}
\langle f | g \rangle = \int_0^{2\pi} \diff \vartheta\,  f(\vartheta)^* g(\vartheta) = \int_0^{2\pi} \diff\vartheta\, 
\langle f | \vartheta \rangle \langle \vartheta | g \rangle ,
\end{align}
 for all $f,g \in L^2[0,2\pi]$. We therefore infer the completeness relation
\begin{align}
\int_0^{2\pi} \diff \vartheta | \vartheta \rangle \langle \vartheta | = \mathbbm{1} .
\end{align}
Furthermore from 
\begin{align}
\mathbbm{1}^2 = \int_0^{2\pi}  \diff \vartheta \int_0^{2\pi} \diff \vartheta_0 \, | \vartheta \rangle \langle \vartheta | \vartheta_0 \rangle \langle \vartheta_0 | \stackrel{!}{=} \mathbbm{1}  ,
\end{align}
we conclude the generalized orthogonality
\begin{align}
\langle \vartheta | \vartheta_0 \rangle = \delta(\vartheta-\vartheta_0) .
\end{align}

We write $\{ | n \rangle : n\in \mathbb{Z} \}$ for the standard orthonormal basis (ONB)  in $\mathcal{H}$ with real-space representation $\langle \vartheta | n \rangle = \exp(i n \vartheta)/ \sqrt{2\pi}$. We define the (non-Hermitian) operator $\mathcal{L}$ in $\mathcal{H}$ via its matrix elements $\langle m| \mathcal{L} n\rangle$ \footnote{Since $\mathcal{L}$ is non-Hermitian one has to carefully indicate whether operators act to the right or the left. }
of the Fokker-Planck operator  [3]
\begin{align}\label{eq:matrix_elements}
\mathcal{L}_{mn} = \langle m | \mathcal{L}  n \rangle = \int_0^{2\pi} \frac{\diff \vartheta}{2\pi} e^{-i m \vartheta } \mathcal{L} e^{in \vartheta} 
= (-D_{\text{rot}} m^2 - i m \omega ) \delta_{mn} + \frac{\gamma}{2} m ( \delta_{m,n+1} - \delta_{m,n-1} ) .
\end{align}
In particular the matrix representation becomes tridiagonal. 
We solve for right and left eigenstates of the Fokker-Planck operator 
\begin{align}
\mathcal{L} | r_\lambda \rangle &= -\lambda | r_\lambda \rangle , \qquad 
 \mathcal{L}^\dagger | l_\lambda \rangle = -\lambda^* | l_\lambda \rangle ,
\end{align}
by expansion in the ONB
\begin{align}
| r_\lambda \rangle = \sum_n  | n \rangle \langle n | r_\lambda \rangle , \qquad 
| l_\lambda \rangle = \sum_n | n \rangle \langle n | l_\lambda \rangle .
\end{align}
Then the expansion coefficients $\langle n | r_\lambda \rangle$ and $\langle n | l_\lambda \rangle $ diagonalize the matrix $\langle m | \mathcal{L}  n \rangle$ 
\begin{align}
 \sum_n \langle m | \mathcal{L}  n\rangle \langle n | r_\lambda \rangle = -\lambda  \langle m | r_\lambda \rangle, \qquad 
  \sum_n \langle m | \mathcal{L}^\dagger  n\rangle \langle n | l_\lambda \rangle = -\lambda  \langle m | l_\lambda \rangle .
\end{align}

In particular, the left and right eigenstates can be chosen to be orthonormal and we assume that they span the entire Hilbert space
\begin{align}
\langle l_\mu | r_\lambda \rangle &= \delta_{\mu\lambda} , \qquad 
\sum_{\lambda} | r_\lambda \rangle \langle l_\lambda | = \mathbbm{1} ,
\end{align}
where $\sum_\lambda$ means that we are summing over all eigenspaces. By conservation of probability the left eigenstate to eigenvalue $0$ is trivial: $\langle l_0 | = \langle 0 |$, see Eq.~\eqref{eq:matrix_elements}. The corresponding right eigenstate $| r_0 \rangle$ is the stationary state with real space representation $\langle \vartheta | r_0 \rangle = p^{\text{st}}(\vartheta)$, where the normalizing factor is chosen such that $\langle l_0 | r_0 \rangle = 1$. 

The angular propagator $P(\vartheta, t| \vartheta_0)$ encoding the probability to find an angle $\vartheta$ at lag time $t$ given the initial orientation at time $t=0$ was $\vartheta_0$ can then be written conveniently in abstract form as
\begin{align} P(\vartheta, t| \vartheta_0 ) &= e^{\mathcal{L} t} \delta(\vartheta-\vartheta_0) = \sum_\lambda e^{-\lambda t} \langle  \vartheta | r_\lambda \rangle \langle l_\lambda |\vartheta_0 \rangle  = \langle \vartheta | \sum_\lambda  e^{\mathcal{L} t} r_\lambda \rangle \langle l_\lambda | \vartheta_0 \rangle = \langle \vartheta | e^{\mathcal{L} t} \vartheta_0 \rangle .
\end{align}

\subsection{Perturbation theory}
The full propagator $\tilde{\mathbb{P}}(\vec{k} \vartheta t | \vartheta_0)$ solves the Fokker-Planck equation
\begin{align}
\partial_t \tilde{\mathbb{P}} = (\mathcal{L} + \delta \mathcal{L}_{\vec{k}} ) \tilde{\mathbb{P}} ,
\end{align}
with 
the perturbation
\begin{align}
\delta \mathcal{L}_{\vec{k}} =& -i \vec{k} \cdot\vec{V} - D_\perp k^2 - (D_\parallel-D_\perp) (\vec{u}^\parallel \cdot \vec{k})^2 .
\nonumber \\
=&   -i k_x v \cos\vartheta - i k_y v \sin\vartheta 
+ i k_x v_1 \sin(\varphi_H + \vartheta) - i k_y v_1 \cos(\vartheta_H + \vartheta) + i k_x v_2 \sin[2 (\varphi_H+ \vartheta)] + i k_y v_2 \cos[2 (\varphi_H+ \vartheta)] \nonumber \\ &  + i v_3 k_y - \frac{D_\parallel-D_\perp}{2} [ (k_x^2- k_y^2 ) \cos(2 (\varphi_H+ \vartheta)) + 2  k_x k_y \sin(2(\varphi_H+ \vartheta)) ] - \frac{D_\parallel+D_\perp}{2} k^2 .
\end{align}
In the following we discard all terms due to translational motion and keep only 
\begin{align}
\delta \mathcal{L}_{\vec{k}} = -i v \vec{k} \cdot\vec{u} = - i k_x v \cos\vartheta - i k_y v \sin\vartheta ,
\end{align}
with corresponding matrix elements
\begin{align}
(\delta \mathcal{L}_{\vec{k}} )_{mn} = \langle m | \delta \mathcal{L}_{\vec{k}}  n \rangle = - \frac{i k_x v}{2} ( \delta_{m,n+1} + \delta_{m,n-1} ) - \frac{ k_y v}{2} ( \delta_{m,n+1} - \delta_{m,n-1} ) .
\end{align}

The formal solution is then given by 
\begin{align}
\tilde{\mathbb{P}}(\vec{k}\vartheta t | \vartheta_0) = e^{(\mathcal{L} +\delta \mathcal{L}_{\vec{k}} ) t} \delta(\vartheta-\vartheta_0) = 
\langle \vartheta | e^{(\mathcal{L} +\delta \mathcal{L}_{\vec{k}} ) t}  \vartheta_0 \rangle .
\end{align}
Then the intermediate scattering function in the stationary state is obtained as
\begin{align}
F(\vec{k},t) &= \int_0^{2\pi} \diff \vartheta  \int_0^{2\pi} \diff \vartheta_0 \langle l_0 | \vartheta \rangle \langle \vartheta | e^{(\mathcal{L} +\delta \mathcal{L}_{\vec{k}} ) t}  \vartheta_0 \rangle \langle \vartheta_0 | r_0 \rangle 
= \langle l_0 |  e^{(\mathcal{L} +\delta \mathcal{L}_{\vec{k}} ) t}  r_0 \rangle .
\end{align}
We use the Dyson representation [4] for the time evolution operator
\begin{align}
e^{(\mathcal{L}+\delta \mathcal{L}_{\vec{k}} ) t } = e^{\mathcal{L} t} + \int_0^t \diff s \, e^{\mathcal{L} (t-s)} \delta \mathcal{L}_{\vec{k}}  e^{(\mathcal{L}+\delta \mathcal{L}_{\vec{k}} ) s }  ,
\end{align}
and expand to second order in the perturbation
\begin{align}
e^{(\mathcal{L}+\delta \mathcal{L}_{\vec{k}} ) t } = e^{\mathcal{L} t} + \int_0^t \diff s\, e^{\mathcal{L} (t-s)} \delta \mathcal{L}_{\vec{k}}  e^{\mathcal{L} s }  + \int_0^t \diff s \int_0^s \diff u \,  e^{\mathcal{L} (t-s)} \delta \mathcal{L}_{\vec{k}} e^{\mathcal{L} (s-u)} \delta \mathcal{L}_{\vec{k}}  e^{\mathcal{L} u } + O(\delta \mathcal{L}_{\vec{k}})^3 ,
\end{align} 
and sandwich it between the states $\langle l_0 | $ and $|r_0 \rangle$. Note further that $\langle l_0 | e^{\mathcal{L} t} = \langle e^{\mathcal{L}^\dagger t} l_0 |  = 
\langle l_0 |$ by conservation of probability and $e^{\mathcal{L} t} | r_0 \rangle = | r_0 \rangle $  by the definition of the stationary state.  This yields 
\begin{align}
F(\vec{k},t) &= 1  +   \int_0^t\! \diff s\,  \langle l_0 | e^{\mathcal{L} (t-s)}  \delta \mathcal{L}_{\vec{k}}   e^{\mathcal{L} s } r_0\rangle + 
\int_0^t \diff s \int_0^s\! \diff u \langle l_0 | e^{\mathcal{L} (t-s)} \delta \mathcal{L}_{\vec{k}} e^{\mathcal{L}_{\vec{k}} (s-u)} \delta \mathcal{L}_{\vec{k}}  e^{\mathcal{L} u }  r_0 \rangle 
 +
 O(|\vec{k}|^3) \nonumber \\
&= 1  +  t \langle l_0 |   \delta \mathcal{L}_{\vec{k}}   r_0\rangle + 
\sum_\lambda\int_0^t\! \diff s \int_0^s\!\diff u \,\langle l_0 |\delta \mathcal{L}_{\vec{k}} e^{\mathcal{L} (s-u)}  r_\lambda \rangle \langle l_\lambda |  \delta \mathcal{L}_{\vec{k}}   r_0 \rangle 
 +
O(|\vec{k}|^3)
\nonumber \\
&= 1  +  t \langle l_0 |   \delta \mathcal{L}_{\vec{k}}  r_0\rangle + 
\sum_\lambda\int_0^t\! \diff s \int_0^s\! \diff u\, e^{-\lambda (s-u)} \langle l_0 |\delta \mathcal{L}_{\vec{k}}   r_\lambda \rangle \langle l_\lambda |  \delta \mathcal{L}_{\vec{k}}   r_0 \rangle 
 +
O(|\vec{k}|^3)
\nonumber \\
&= 1  +  t \langle l_0 |   \delta \mathcal{L}_{\vec{k}}  r_0\rangle + 
\sum_\lambda \frac{e^{-\lambda t} + \lambda t -1}{\lambda^2} 
 \langle l_0 |\delta \mathcal{L}_{\vec{k}}   r_\lambda \rangle \langle l_\lambda |  \delta \mathcal{L}_{\vec{k}}   r_0 \rangle 
 +
O(|\vec{k}|^3) .
\end{align}  
For the cumulant-generating function  this implies
\begin{align}
\ln F(\vec{k}, t) &= t \langle l_0 | \delta \mathcal{L}_{\vec{k}} r_0 \rangle + \sum_\lambda \frac{e^{-\lambda t} + \lambda t -1}{\lambda^2} 
 \langle l_0 |\delta \mathcal{L}_{\vec{k}}   r_\lambda \rangle 
\langle l_\lambda |  \delta \mathcal{L}_{\vec{k}}   r_0 \rangle 
- \frac{t^2}{2} \langle l_0 | \delta \mathcal{L}_{\vec{k}}  r_0 \rangle \langle l_0 | \delta \mathcal{L}_{\vec{k}}  r_0 \rangle + O(|\vec{k}|^3) \nonumber \\
 &= t \langle l_0 | \delta \mathcal{L}_{\vec{k}} r_0 \rangle + \sum_{\lambda\neq 0} \frac{e^{-\lambda t} + \lambda t -1}{\lambda^2} 
 \langle l_0 |\delta \mathcal{L}_{\vec{k}}   r_\lambda \rangle \langle l_\lambda |  \delta \mathcal{L}_{\vec{k}}   r_0 \rangle  + O(|\vec{k}|^3) . 
\end{align}
On the other hand the cumulant expansion yields
\begin{align}
\ln F(\vec{k}, t) &= \ln \langle \exp(- i \vec{k} \cdot \Delta \vec{r}(t) ) \rangle = \ln \left\langle 1 - i \vec{k} \cdot \Delta \vec{r}(t) - \frac{1}{2} [\vec{k} \cdot \Delta \vec{r}(t)  ]^2 + O(|\vec{k}|^3) \right\rangle \nonumber \\
&= - i \vec{k} \cdot \langle \Delta \vec{r}(t) \rangle - \frac{1}{2} \left\{ \langle [ \vec{k} \cdot \Delta \vec{r}(t) ]^2 \rangle  -  [ \langle \vec{k} \cdot \Delta \vec{r}(t) \rangle]^2  \right\}  .
\end{align}
In particular, we read off the mean drift velocity along the direction $\vec{n} := \vec{k} /k $ 
\begin{align}
\vec{n} \cdot  \frac{\diff}{\diff t} \langle \Delta \vec{r}(t) \rangle  = \frac{i}{k} \langle l_0 | \delta \mathcal{L}_{\vec{k}}   r_0 \rangle ,
\end{align}
as well as the variance
\begin{align}
\textsf{Var}[\vec{n}\cdot \Delta \vec{r}(t)] &= \langle [ \vec{n} \cdot \Delta \vec{r}(t) ]^2 \rangle  -  [ \langle \vec{n} \cdot \Delta \vec{r}(t) \rangle]^2 \nonumber \\
&= \frac{2}{k^2}  \sum_{\lambda\neq 0} \frac{ 1- \lambda t- e^{-\lambda t}}{\lambda^2} 
 \langle l_0 |\delta \mathcal{L}_{\vec{k}}   r_\lambda \rangle \langle l_\lambda |  \delta \mathcal{L}_{\vec{k}}   r_0 \rangle  .
\end{align}\label{eq:Diffusion_analytics_SM}
In particular for the corresponding diffusion coefficient
\begin{align}
D_{\vec{n}} := \lim_{t\to \infty} \frac{1}{2} \frac{\diff}{\diff t} \textsf{Var}[\vec{n}\cdot \Delta \vec{r}(t)]  = \frac{-1}{k^2}  \sum_{\lambda\neq 0} \frac{1}{\lambda} 
 \langle l_0 |\delta \mathcal{L}_{\vec{k}}   r_\lambda \rangle \langle l_\lambda |  \delta \mathcal{L}_{\vec{k}}   r_0 \rangle .
\end{align}
For short times the variance grows as $t^2$ since
\begin{align}
\textsf{Var}[\vec{n}\cdot \Delta \vec{r}(t)] &= \frac{1}{k^2} t^2  \sum_{\lambda\neq 0}  
 \langle l_0 |\delta \mathcal{L}_{\vec{k}}   r_\lambda \rangle \langle l_\lambda |  \delta \mathcal{L}_{\vec{k}}   r_0 \rangle + O(t^3) \nonumber \\
 &=  \frac{1}{k^2} t^2    
 \langle l_0 |\delta \mathcal{L}_{\vec{k}}    \delta \mathcal{L}_{\vec{k}}   r_0 \rangle 
-  \frac{1}{k^2} t^2    
 \langle l_0 |\delta \mathcal{L}_{\vec{k}}   r_0 \rangle \langle l_0  \delta \mathcal{L}_{\vec{k}}   r_0 \rangle 
+ O(t^3) .
\end{align}

\subsection{Numerical results for the eigenspectrum}


\begin{wrapfigure}{r}{.5\linewidth}
\includegraphics[width=\linewidth]{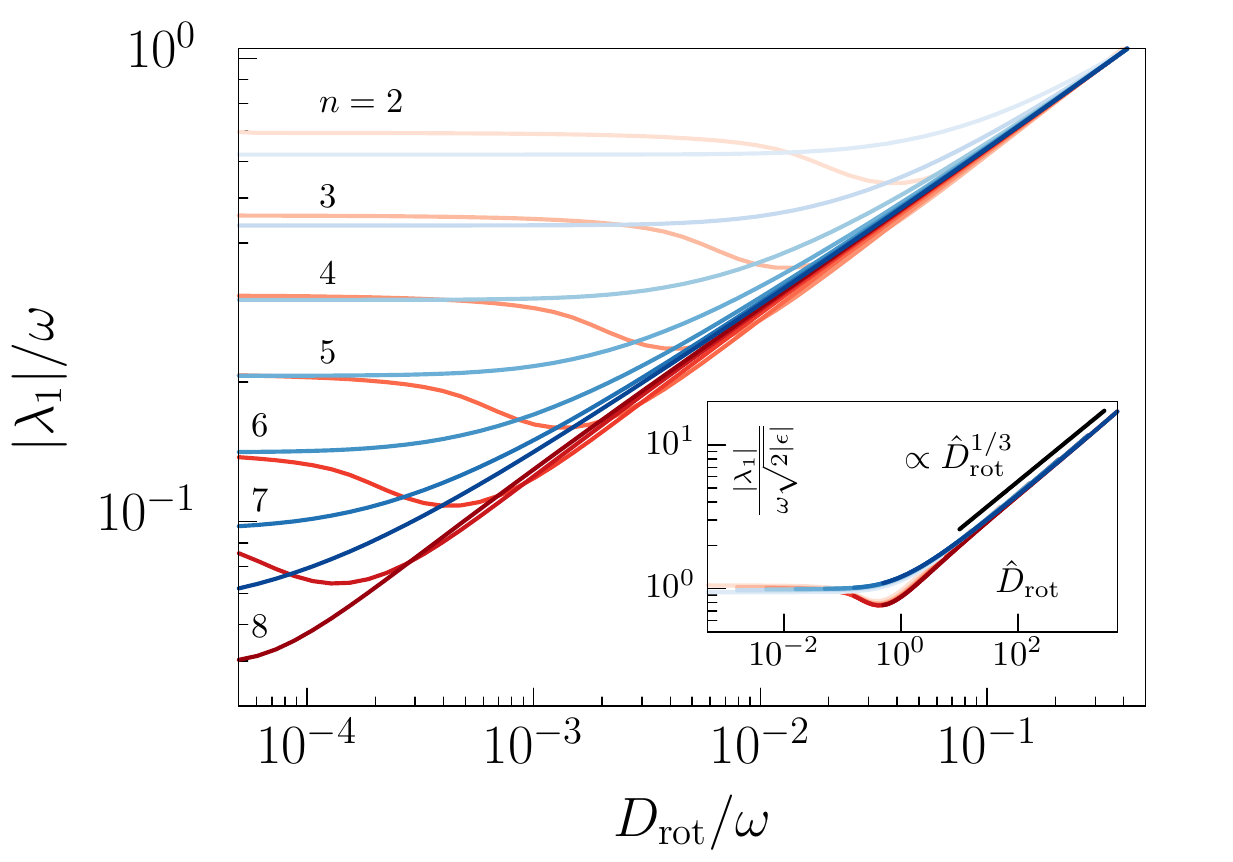}%
\caption{Absolute value of the first non-vanishing eigenvalue, $|\lambda_1|$, vs. 
diffusivity $D_{\text{rot}}$ for decreasing separation parameter 
$\epsilon = (\gamma-\omega)/\omega = \pm 10^{-n/3}, n=2,\ldots,8$. 
The red lines are for $\epsilon> 0$, while blue ones correspond to $\epsilon < 0$. The color becomes darker as the absolute value of  the separation parameter approaches zero. 
Inset: Rescaled eigenvalues $|\lambda_1| / \omega \sqrt{2 |\epsilon|}$ vs. reduced rotational diffusion coefficient $\hat{D}_{\text{rot}} := |\epsilon|^{-3/2} D_{\text{rot}}/\omega$. 
The straight solid line indicates a power law $\propto \hat{D}_{\text{rot}}^{1/3}$ and serves as guide to the eye. 
}
 \label{fig:FigS2}
\end{wrapfigure}
Numerical results for the magnitude of the first nontrivial eigenvalue $\lambda_1$ for decreasing noise are shown in Fig.~\ref{fig:FigS2} on double-logarithmic scales. For $D_{\text{rot}} \to 0$ they approach  finite values which become smaller and smaller as the separation parameter $\epsilon = (\gamma-\omega)/\omega $ decreases. Empirically we find that $\lambda_n \propto \sqrt{\epsilon}$ in the limit of vanishing noise. 
The shapes of the curves above the bifurcation  become identical as the bifurcation is approached, 
$\epsilon\to 0$, and similarly below the bifurcation. This observation holds generally for the low-lying eigenvalues $\lambda_n, n=1,2,\ldots$. The curves can be superimposed upon shifting vertically by their low-noise values and proper rescaling of the rotational diffusion constant. We achieve data collapse using the reduced rotational diffusion coefficient $\hat{D}_{\text{rot}} \propto |\epsilon|^{-3/2} D_{\text{rot}} $ suggesting the scaling law 
\begin{align}\label{eq:scaling_law}
\lambda_n /\omega= \sqrt{2 |\epsilon| }\Lambda_{n,\pm}(\hat{D}_{\text{rot}} ) ,
\end{align}  
with dimensionless scaling functions $\Lambda_{n,\pm}(.)$ for $\epsilon \gtrless 0$ (it turns out that the factor $2$ in the square root renders $\Lambda_{n,+}(\hat{D}_{\text{rot}}\to 0) = n$).    

The inset of Fig.~\ref{fig:FigS2} exhibits the scaling behavior in the vicinity of the bifurcation for small noise.
By construction  both scaling functions saturate for small reduced rotational diffusion coefficient $\hat{D}_{\text{rot}} \ll 1 $, while for large $\hat{D}_{\text{rot}} \gg 1 $ they behave as a power law with an exponent of $1/3$: 
\begin{align}
\Lambda_{n,\pm}(\hat{D}_{\text{rot}}) \sim C_n  \hat{D}_{\text{rot}}^{1/3} \qquad \text{for } \hat{D}_{\text{rot}} \to \infty ,
\end{align}
with constants $C_n$ independent of the sign of $\epsilon$. In the next subsection the scaling behavior of the low-noise limit as well as the proper choice of $\hat{D}_{\text{rot}}$ is rationalized in terms of a 
harmonic-oscillator picture. 
\subsection{Harmonic approximation}

\begin{wrapfigure}{r}{.35\linewidth}
\centering
a) \includegraphics[width=\linewidth,keepaspectratio=true]{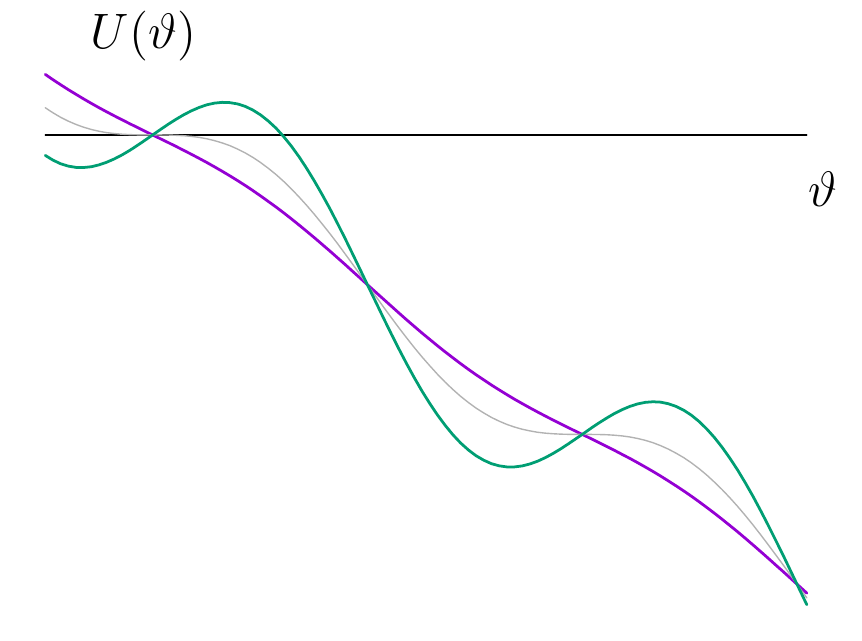}
b) \includegraphics[width=\linewidth,keepaspectratio=true]{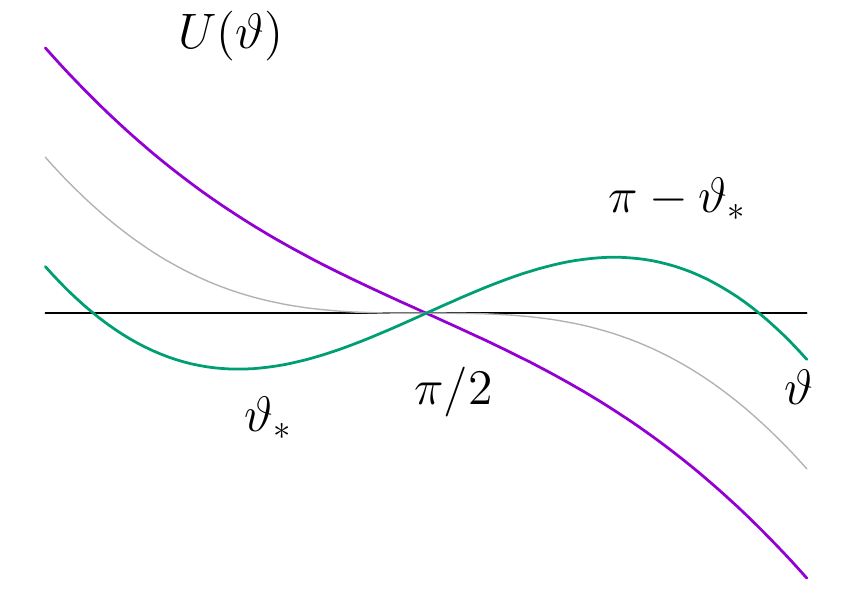}
\caption{a) Titled washboard potential $U(\vartheta)$. b) Zoom close to the classical bifurcation.  }
\label{fig:washboard}
\end{wrapfigure}
We rewrite the stochastic e.o.m. in terms of an effective potential $U(\vartheta)$
\begin{align}
\dot{\vartheta} = - \frac{D_{\text{rot}}}{k_B T}  \partial_\vartheta U + \zeta(t)  .
\end{align}
Here $D_{\text{rot}}/k_B T$ is then the mobility for the overdamped motion of the angle and the potential reads 
\begin{align}
 \frac{D_{\text{rot}}}{k_B T} U(\vartheta) = - \omega \vartheta - \gamma \cos \vartheta. 
\end{align}
For $\gamma \geq \omega$ the
potential displays a minimum at  $\vartheta_* \in [0,\pi/2]$ with $\sin \gamma_* = \omega/\gamma$
while a maximum is displayed at $\pi- \vartheta_* \in [\pi/2, \pi]$, see Fig.~\ref{fig:washboard} for illustration.  
Let's make a harmonic approximation for the potential close to the minimum, 
\begin{align}
\frac{D_{\text{rot}}}{k_B T} U(\vartheta)  
&=\frac{D_{\text{rot}}}{k_B T}  U(\vartheta_*) + \frac{( \vartheta-\vartheta_*)^2}{2} \gamma \cos \vartheta_* + O(\vartheta-\vartheta_*)^3 .
\end{align}
For the stochastic e.o.m.\@ this results in a linear Langevin equation
\begin{align}
\dot{\vartheta} &= -\frac{1}{\tau} (\vartheta-\vartheta_*) + \zeta(t) ,
\end{align}
with relaxation rate
\begin{align}
\frac{1}{\tau} = \gamma \cos\vartheta_* = \sqrt{\gamma^2-\omega^2}.
\end{align}
The eigenvalues of the overdamped harmonic oscillator are simply
\begin{align}
\lambda_n = \frac{1}{\tau} n.
\end{align}
The picture of the harmonic oscillator is expected to be valid for the low eigenvalues if the escape rate for diffusing over the barrier is small compared to the relaxation rate. For large barriers $\Delta U := U(\pi-\vartheta_*)- U(\vartheta_*)$ the escape is exponentially suppressed. $\propto\exp(- \Delta U/k_B T)$, by Kramers' law, correspondingly we require that the barrier is high with respect to the thermal energy
\begin{align}
1 \ll \frac{\Delta U}{k_B T} &=   \frac{1}{k_B T } [ U(\pi-\vartheta_*) - U(\vartheta_*) ] 
\nonumber \\
&= \frac{1}{D_{\text{rot}}} [ 2 \omega \arcsin (\omega/\gamma) - \pi \omega + 2 \sqrt{\gamma^2-\omega^2} ] .
\end{align} 
The condition is certainly fulfilled for $D_{\text{rot}} \to 0$ for fixed $\gamma>\omega$. Here we want to investigate the scaling limit of $D_{\text{rot}}/\omega \to 0$ and $\epsilon := (\gamma-\omega)/\omega \to 0$. Then upon series expansion the condition of high barrier translates to 
\begin{align}
1 \ll \frac{\Delta U}{k_B T} = \frac{4 \sqrt{2}}{3} \frac{\omega}{D_{\text{rot}}} \epsilon^{3/2} + O(\ldots) .
\end{align}
This is the condition, Eq.~(14), of the main text.

\subsection*{Matrix elements of the harmonic oscillator}


For small $D_{\text{rot}}$ and $\gamma> \omega$ we perform the harmonic oscillator approximation around the minimum $\vartheta_*$,  $x :=\vartheta - \vartheta_*$ 
\begin{align}
\mathcal{L} &= D_{\text{rot}} \frac{\partial^2 }{\partial \vartheta^2} - \frac{\partial}{\partial \vartheta} ( \omega- \gamma \sin\vartheta) = D_{\text{rot}} \frac{\partial^2 }{\partial x ^2} - \frac{\partial}{\partial x} [ \omega- \gamma \sin(\vartheta_* + x) ] \nonumber \\
&\approx D_{\text{rot}} \frac{\partial^2 }{\partial x^2} - \frac{\partial}{\partial x} ( \omega- \gamma \sin\vartheta_* - x \gamma \cos\vartheta_* ) .
\end{align} 
Since $\gamma \sin \vartheta_* = \omega$ and $\gamma \cos\vartheta_* = \sqrt{\gamma^2-\omega^2 }=: 1/\tau$ this simplifies to 
\begin{align}
\mathcal{L} = D_{\text{rot}} \frac{\partial^2 }{\partial x^2} + \frac{\partial}{\partial x} \left(  \frac{x}{\tau}  \right) .
\end{align} 
This Fokker-Planck operator corresponds to the effective potential 
\begin{align}
\frac{D_{\text{rot}} U(x)}{k_B T} = \frac{x^2}{2 \tau } .
\end{align}
The angular oscillator width $\langle (\vartheta-\vartheta_*)^2 \rangle^{1/2} = \sqrt{\langle x^2 \rangle} = \sqrt{D_{\text{rot}} \tau}$ quantifies the mean-square fluctuations. For $D_{\text{rot}}\to 0$ the eigenfunctions become more and more localized.  

For the perturbation we expand also to linear order
\begin{align}
\delta \mathcal{L}_{\vec{k}} &= - i k_x v \cos \vartheta - i k_y \sin \vartheta 
= - i k_x v \cos (\vartheta_*+ x) - i k_y v \sin(\vartheta_* + x) \nonumber \\
&=  -i k_x v \cos \vartheta_* - i k_y v \sin\vartheta_* + ( i k_x v  \sin\vartheta_* - i k_y v  \cos\vartheta_*) x + O(x^2) .   
\end{align}
Thus to leading order only the diagonal components are non-vanishing
\begin{align}
\langle l_\mu | \delta \mathcal{L}_{\vec{k}} | r_\lambda \rangle = (-i k_x v \cos \vartheta_* - i k_y v \sin\vartheta_*) \delta_{\mu\lambda} + O(.) .
\end{align}
This yields for the mean drift 
\begin{align}
\vec{n} \cdot  \frac{\diff}{\diff t} \langle \Delta \vec{r}(t) \rangle  = \frac{i}{k} \langle l_0 | \delta \mathcal{L}_{\vec{k}}  | r_0 \rangle \to (n_x v \cos \vartheta_* +n_y v \sin\vartheta_*) , 
\end{align}
with $\vec{n} = \vec{k}/k$. This reproduces the classical result. 

The variance $\textsf{Var}_{\vec{n}}[\Delta \vec{r}(t)$ vanishes as $D_{\text{rot}}\to 0$. To obtain the leading order we need to evaluate the matrix elements $\langle l_\mu | x | r_\lambda \rangle$ for $\mu\neq \lambda$. 
The fastest way is to use the gauge transformation to map the Fokker-Planck operator to a Schr\"odinger Hamiltonian [3]
 Define the Hamiltonian by the gauge transform
\begin{align}
\mathcal{H} &= -e^{  U(x)/2 k_B T  } \mathcal{L} e^{  -U(x)/2 k_B T  } =  -e^{  x^2 /4 D_{\text{rot}} \tau  } \mathcal{L} e^{ - x^2 /4 D_{\text{rot}} \tau  } \nonumber \\
&= - D_{\text{rot}} \frac{\partial^2 }{\partial x^2} + \frac{x^2}{4 D_{\text{rot}} \tau^2}  - \frac{1}{2\tau}, 
\end{align}
which is Hermitian in $L^2(\mathbb{R})$. Adjoining yields
\begin{align}
\mathcal{H} &= -e^{- U(x)/2 k_B T  } \mathcal{L}^\dagger e^{  U(x)/2 k_B T  }.
\end{align}
We substitute $D_{\text{rot}} \mapsto \hbar^2 /2 m, 1/2 D_{\text{rot}} \tau^2 \mapsto m \omega^2 $  and find the shifted harmonic oscillator 
\begin{align}
\mathcal{H} = \frac{\hbar^2}{2m}\frac{\partial^2}{\partial x^2} + \frac{m \omega^2}{2} x^2 - \frac{\hbar \omega}{2} ,
\end{align} 
with eigenvalues $\lambda_n = n \hbar \omega = n/\tau$. In particular the quantum oscillator length $\sqrt{\hbar/2 m \omega} $ maps to the angular oscillator width $\sqrt{D_{\text{rot}} \tau}$. The matrix elements of $x$ are thus expected to be of order $\sqrt{D_{\text{rot}} \tau}$

The corresponding mapping for the eigenfunctions is then 
\begin{align}
r_n(x) &= e^{- U(x)/2 k_B T} \psi_n(x) ,\\
l_n(x) &= e^{ U(x)/2 k_B T} \psi_n(x) , 
\end{align}
with $\psi_n(x)$ the eigenfunctions of $\mathcal{H}$. The matrix elements of an operator $\mathcal{V}$ are obtained as
\begin{align}
\langle l_m | \mathcal{V}  r_n\rangle = \int \diff x \,l_m(x)^*\, \mathcal{V} \, r_n(x) = \int \diff x\, \psi_m(x)^*  [ e^{U(x)/2k_B T} \mathcal{V} e^{-U(x)/2k_B T} ] \psi_n(x)   .
\end{align}
For the case of $\mathcal{V} = x$ the exponentials cancel, and we read off (with harmonic oscillator creation and annihilation operators $a, a^\dagger$) 
\begin{align}
\langle l_m | x | r_n \rangle &= \langle \psi_m | x | \psi_n \rangle =   \langle \psi_m | \sqrt{\frac{\hbar}{2m \omega}} \left( a^\dagger + a\right) | \psi_n \rangle \nonumber \\
 &=  \sqrt{\frac{\hbar}{2m \omega}}
 \left( \delta_{m,n+1} \sqrt{n+1} + \delta_{m,n-1} \sqrt{n} \right) 
\mapsto \sqrt{D_{\text{rot}} \tau}  \left( \delta_{m,n+1} \sqrt{n+1} + \delta_{m,n-1} \sqrt{n} \right) . 
\end{align}
We thus read off the relevant matrix elements for the variance/diffusion coefficient
\begin{align}
\langle l_0 | \delta \mathcal{L}_{\vec{k}}  r_\lambda \rangle &=
\langle l_\lambda | \delta \mathcal{L}_{\vec{k}}  r_0 \rangle =  i  v \sqrt{D_{\text{rot}} \tau}  (  k_x   \sin\vartheta_* -  k_y  \cos\vartheta_*) \delta_{\lambda,1} . 
\end{align}
The picture of the harmonic oscillator is valid only for $\hat{D}_{\text{rot}} \ll 1$. In this regime the formulas for the variance and diffusion simplify since only a single term contributes
\begin{align}
\textsf{Var}[\vec{n}\cdot \Delta \vec{r}(t)] 
&= \frac{2}{k^2}  \sum_{\lambda\neq 0} \frac{ 1- \lambda t- e^{-\lambda t}}{\lambda^2} 
 \langle l_0 |\delta \mathcal{L}_{\vec{k}}   r_\lambda \rangle \langle l_\lambda |  \delta \mathcal{L}_{\vec{k}}   r_0 \rangle \nonumber \\
 &= 2 \left( \frac{t}{\tau} -1+  e^{-t/\tau} \right)   (v \tau)^2  D_{\text{rot}} \tau (n_x \sin \vartheta_* - n_y \cos\vartheta_*)^2 ,
 \end{align}
and infer the associated long-time diffusion coefficient
\begin{align}
D_{\vec{n}}   &=   (v \tau)^2  D_{\text{rot}}  (n_x \sin \vartheta_* - n_y \cos\vartheta_*)^2 .
\end{align}
In particular, for small $D_{\text{rot}} \ll \omega$, $D_0 \approx v^2 D_{\text{rot}}/2\omega^2$ in the harmonic oscillator approximation 
\begin{align*}
D_{\vec{n}}/D_0  = 2 (\omega \tau)^2  (n_x \sin \vartheta_* - n_y \cos\vartheta_*)^2
\approx \frac{1}{\epsilon} (n_x \sin \vartheta_* - n_y \cos\vartheta_*)^2 .
\end{align*}
This result yields prefactors to the scaling law derived in the main text. 

\section{Supplementary figure on mean drift velocity} 

The mean drift velocity
is evaluated directly from the stationary distribution 
\begin{align}
v \begin{pmatrix}
\langle \cos \vartheta \rangle \\
\langle \sin \vartheta \rangle 
\end{pmatrix} 
= \int_0^{2\pi}  v \begin{pmatrix}
 \cos \vartheta  \\
 \sin \vartheta 
\end{pmatrix} 
p^{\text{st}}(\vartheta) \diff \vartheta .
\end{align}

\begin{figure}[!h]
\centering
\includegraphics[width=0.9\linewidth]{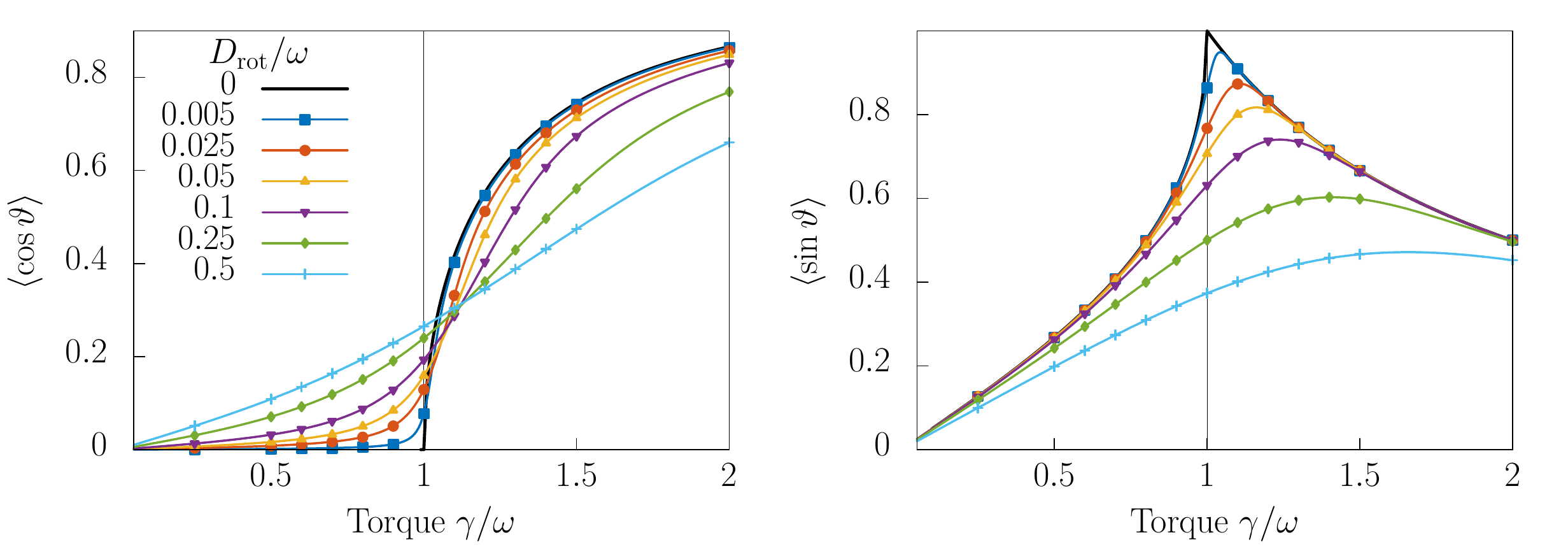}%
\caption{Mean drift velocity in the stationary state  along the horizontal (left) and vertically against the field (right) as a function of the torque $\gamma$ for various noise strength $D_{\text{rot}}$. Symbols correspond to simulation, full lines to numerical results. 
}
\label{fig:Mean_drift}
\end{figure}

The stationary distribution is known explicitly [3]
\begin{align}
 p^{\text{st}}(\vartheta)  = & 
  \mathcal{N}   e^{-U(\vartheta)/k_B T}  \left[  \int_\vartheta^{2\pi} 
e^{U(\bar{\vartheta})/k_B T} \diff \bar{\vartheta} +
 e^{[U(0)-U(2\pi)]/k_B T}  
\int_0^\vartheta e^{U(\bar{\vartheta})/k_B T } \diff \bar{\vartheta} \right] ,
\end{align}
with the potential
\begin{align}
\frac{D_{\text{rot}} U(\vartheta)}{k_B T} = - \omega \vartheta - \gamma \cos\vartheta .
\end{align}
The normalization factor $\mathcal{N}$ is fixed by imposing $\int_0^{2\pi} p^{\text{st}}(\vartheta)\diff \vartheta= 1$. 

Numerical results for the average sine and cosine are displayed in Fig.~\ref{fig:Mean_drift} but similar figures can also  be found in Ref. [3].

\section{Supplementary figure on influence of translational diffusion on the resonance} 

\begin{figure}[!h]
\centering
\includegraphics[width=0.45\linewidth]{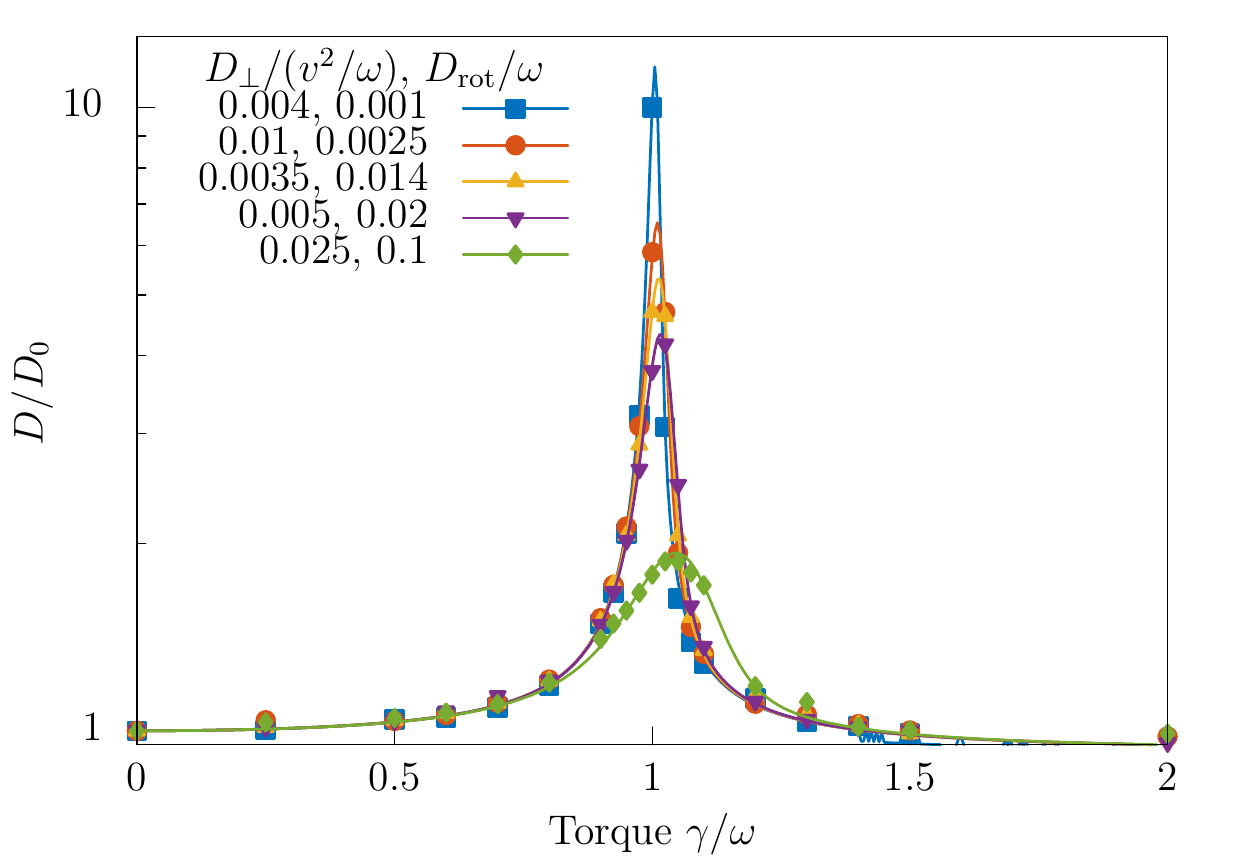}%
\caption{Diffusion coefficient $D$ as a function of the torque $\gamma$ for decreasing orientational diffusion coefficients $D_{\text{rot}}$  and translational diffusion coefficients $D_\perp$ ($D_\parallel=2 D_\perp$).  Symbols correspond to simulation, full lines to numerical results. The values of the perpendicular component of the translational diffusivity $D_\perp$ and rotational diffusivity $D_\mathrm{rot}$ are specified in the legend.
The parallel component always is $D_\parallel=2 D_\perp$, for simplicity $\varphi_H=0$, and we choose parameters such that  $v_1= v_2 = v_3/3 = v$.  
 }
\label{fig:FigS5}
\end{figure}

The techniques elaborated in the main text can be readily extended to include the additional drift terms and the translational diffusion in Eq.~\eqref{eq:Langevin_full}. The resonance for the full model emerges in the simultaneous limit of vanishing separation parameter and all noise terms approaching zero $D_{\text{rot}} \to 0, D_\perp \to 0, D_\parallel \to 0$ such that the ratios $D_\perp/D_{\text{rot}}, D_\parallel/D_{\text{rot}}$ remain fixed. Simulation as well as numerical results are displayed 
 in Fig.~\ref{fig:FigS5}. Although diffusion broadens the resonance, the general picture remains valid.

\section*{Bibliography}

\begin{enumerate}[label={[\arabic*]}]
\item B. ten Hagen, F. K\"{u}mmel, R. Wittkowski, D. Takagi, H. L\"{o}wen, and C. Bechinger, Gravitaxis of asymmetric self-propelled colloidal particles, Nature Communications 5, 4829 (2014).
\item The use of the hydrodynamic expressions for active propulsion is controversial, see Ref. [6].
\item H. Risken, The Fokker-Planck Equation (Springer Berlin Heidelberg, 1989).
\item J. Sakurai and J. Napolitano, Modern quantum mechanics. San Fransico (CA: Addison-Wesley, 2011).
\item Since $\mathcal{L}$ is non-Hermitian one has to carefully indicate whether operators act to the right or the left.
\item B. U. Felderhof, Comment on ``Circular motion of asymmetric self-propelling particles'', Physical Review Letters 113 (2014);
F. K\"{u}mmel, B. ten Hagen, R. Wittkowski, D. Takagi, I. Buttinoni, R. Eichhorn, G. Volpe, H. L\"{o}wen, and C. Bechinger, K\"{u}mmel et al. Reply:, Physical Review Letters 113, 029802 (2014).

\end{enumerate}




\begin{thebibliography}{71}
\bibitem [{foo()}]{footnote1}%
  \BibitemOpen
  \href@noop {} {}\bibinfo {note} {This will somewhat change upon reinstating
  the terms due to translational diffusion which we ignore here.}\BibitemShut
  {Stop}%
\end{thebibliography}
\end{document}